\documentclass[11pt,twoside]{article}

\usepackage{asp2014}

\aspSuppressVolSlug
\resetcounters

\begin{document}

\title{The Observers’ Data Access Portal at the Keck Observatory Archive}


\author{T.~ Oluyide $^1$, M. ~S.~ Lynn $^2$, T.~Coda $^2$, G.~ B. ~Berriman $^2$, M. ~Brown $^2$, L. ~Fuhrman $^2$, C.~ Gelino $^2$, J.~ Good $^2$, J.~ Hayashi $^3$,  C.-H. ~Lee$^3$, J. ~Mader $^3$, M.~ A.~ Swain $^2$}
\affil{$^1$, Caltech/IPAC-NExScI, Pasadena, CA 91125, USA; \email{toluyide@ipac.caltech.edu}}
\affil{$^2$ Caltech/IPAC-NExScI, Pasadena, CA 91125, USA} 
\affil{$^3$ W. M. Keck Observatory, Kamuela, HI 96743, USA}

\paperauthor{T.~ Oluyide}{toluyide@ipac.caltech.edu}{ }{Caltech}{IPAC/NExScI}{Pasadena}{CA}{91125}{USA}
\paperauthor{M. ~S.~Lynn}{mlynn@ipac.caltech.edu}{ }{Caltech}{IPAC/NExScI}{Pasadena}{CA}{91125}{USA}

\paperauthor{T.~ Coda}{tcoda@keck.hawaii.edu}{ }{W. M. Keck Observatory}{ }{Kamuela}{HI}{96743}{USA}
\paperauthor{G.~Bruce Berriman}{gbb@ipac.caltech.edu}{0000-0001-8388-534X}{Caltech}{IPAC/NExScI}{Pasadena}{CA}{91125}{USA}
\paperauthor{G.~Bruce Berriman}{gbb@ipac.caltech.edu}{0000-0001-8388-534X}{Caltech}{IPAC/NExScI}{Pasadena}{CA}{91125}{USA}
\paperauthor{M. ~Brown}{mbrown@keck.hawaii.edu}{ }{W. M. Keck Observatory}{ }{Kamuela}{HI}{96743}{USA}
\paperauthor{L.~ Fuhrman}{lfuhrman@keck.hawaii.edu}{ }{W. M. Keck Observatory}{ }{Kamuela}{HI}{96743}{USA}
\paperauthor{C.~ R.~ Gelino}{cgelino@ipac.caltech.edu}{0000-0001-5072-4574}{Caltech}{IPAC/NExScI}{Pasadena}{CA}{91125}{USA}
\paperauthor{J.~C.~Good}{jcg@ipac.caltech.edu}{ }{Caltech}{IPAC/NExScI}{Pasadena}{CA}{91125}{USA}
\paperauthor{J.~Hayashi}{jhayashi@keck.hawaii.edu}{ }{W. M. Keck Observatory}{ }{Kamuela}{HI}{96743}{USA}
\paperauthor{C.~-H. ~Lee}{clee@keck.hawaii.edu}{ }{W. M. Keck Observatory}{ }{Kamuela}{HI}{96743}{USA}
\paperauthor{J.~ Mader}{jmader@keck.hawaii.edu}{ }{W. M. Keck Observatory}{ }{Kamuela}{HI}{96743}{USA}
\paperauthor{M.~A.~Swain}{mswain@ipac.caltech.edu}{0000-0003-4557-1192}{Caltech}{IPAC/NExScI}{Pasadena}{CA}{91125}{USA}



\begin{abstract}
For all active instruments, the Keck Observatory Archive (KOA) 
 now ingests raw data from the Keck Telescopes within 1 minute of acquisition, quick-look reduced data within 5 minutes of creation, and science-ready reduced data for four instruments as they are created by their automated pipelines. On August 1, 2023, KOA released the Observers’+ Data Access Portal (ODAP), which enables observers at the telescope and their collaborators anywhere in the world to securely monitor and download science, calibration, and quick-look data as they are ingested into the archive. The portal is built using Python Socket.IO WebSockets that ensure metadata appear in the portal as the data themselves are ingested. The portal itself is a dynamic web interface built with React. It enables users to view and customize metadata fields, filter metadata according to data type, and download data as they are ingested or in bulk through wget scripts. Observers have used the ODAP since its release and have provided feedback that will guide future releases.
\end{abstract}



\section{Introduction}
The W. M. Keck Observatory is undergoing a major upgrade of its operational model, in which observational planning, data acquisition, data reduction, and data archiving are tightly coupled. This overall activity is called the Data Services Initiative (DSI) \citep{2022SPIE12186E..0HB}. One component of DSI, the tight coupling of the observatory operations and the Keck Observatory Archive, has recently been completed. Newly acquired raw data are ingested into the archive at NExScI, generally within 1 minute of acquisition, quick-look reduced data are ingested 5 minutes after acquisition of the raw data, and science-ready reduced data are ingested as they are created. A purpose-built web interface, the Observers' Data Acquisition Portal (ODAP), has been deployed to enable observers and their worldwide collaborators to monitor their data as they are archived and download them on demand. Thus new data can arrive on an observer's computer within minutes of acquisition. \cite{2022arXiv221202576B} describes the real-time data acquisition;   this paper describes the portal and its design. 

\section{Operation of the Observers' Data Access Portal}

The PI logs in to the portal with credentials assigned to them by KOA on a semester-by-semester basis for the particular observing program they are executing. They may add collaborators via a web form accessible from the portal. Thus, all collaborators worldwide will have immediate access to that night's data. The portal provides access to each night's data rather than an entire observing run: previous nights' data are only accessible through the KOA landing page \footnote{\url{https://koa.ipac.caltech.edu}}. A "night" in the context of the ODAP is defined as the 24-hour period beginning at 2 p.m. Hawaiian Standard Time (HST). Calibration files acquired at twilight are immediately public and accessible to the observer through the portal, along with science data and night-time calibration data, to which the PI and collaborators have exclusive access for 18 months (12 months for NASA-sponsored data). Quick-look reduced files and science-ready reduced data are made available as they are created until the observing night ends. In particular, any science files created after the end of the night will only be available through the KOA landing page. Such may be the case with especially data-intensive reductions for some instruments.  

Figure 1 shows the ODAP during a typical night's observations. The portal shows the active observing programs to which the user has been granted access; generally, this will include the night's program, plus calibration files (designated by KOA as ENG, or engineering data). The portal shows a table of metadata for all the observations, including filename and data type (science or calibration). Users may configure this table as they wish. As new files are ingested, they are listed in the table in reverse chronological order. The table may be filtered according to needs, for example, to show only science or calibration data.)

\articlefigure{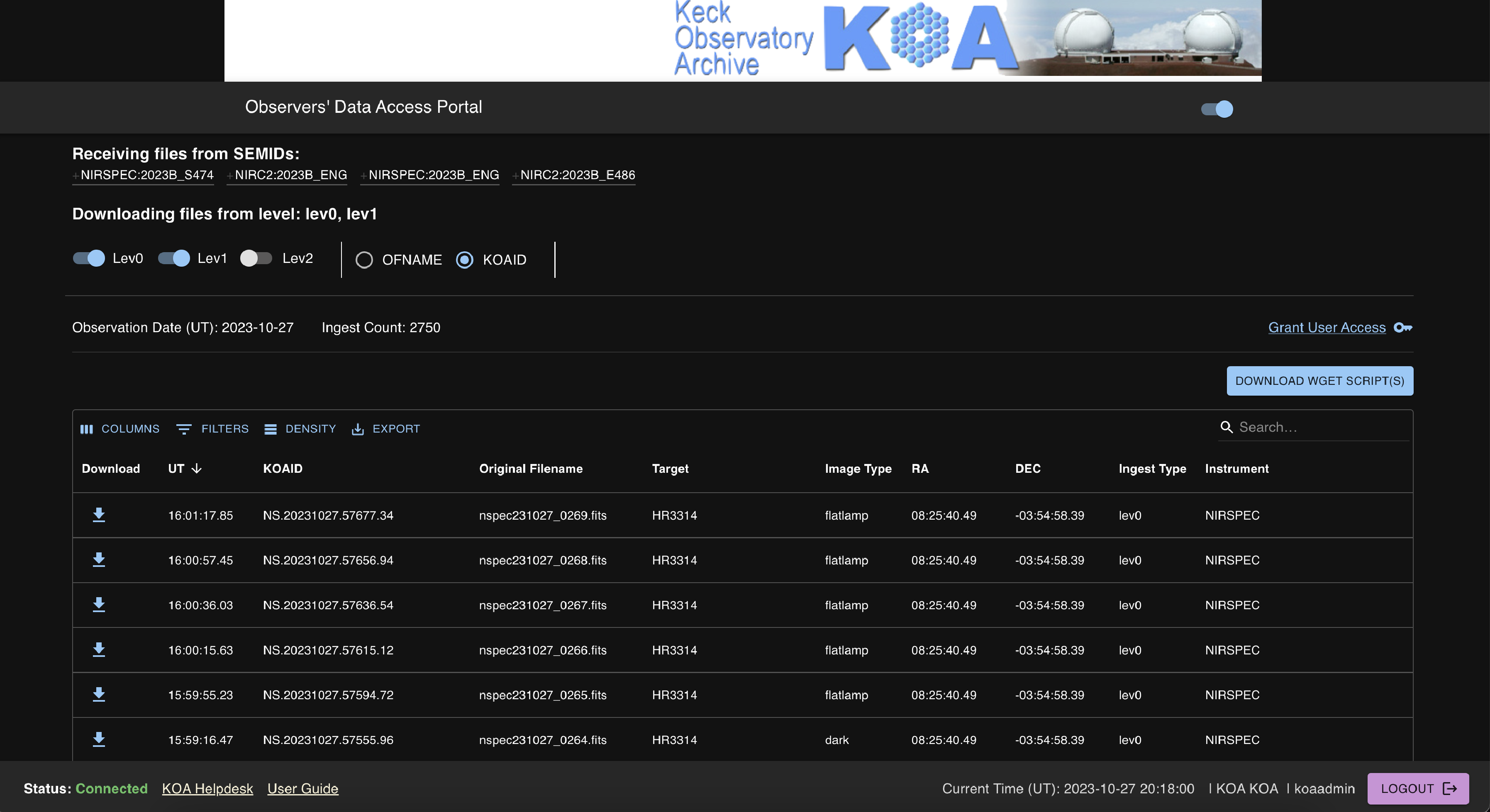}{ex_fig1}{The ODAP during nightly observations}

Various options are provided for data access:
\begin{itemize}
\item Download manually file by file.
\item Download automatically as files are ingested.
\item  Download in bulk via wget scripts (created extemporaneously as soon as the wget option button is pressed). These scripts are usually run at the end of the night, but can be run anytime.
\item Download combinations of raw, quick-look or science-grade files.
\end{itemize}

\section{The Design}

The essential features of the design are:

\begin{itemize}
    \item Use of web sockets, the industry standard technology for communication between two networked clients. It is especially valuable for real-time communication for data that need to update quickly with little latency.
        \item Use of "rooms" to allow the transfer of data to authorized users.
    \item A modern React front-end interface that allows authenticated users to select, filter, and download data.
\end{itemize}

 Figure 2 shows the design components of the ODAP, with the data flowing from the archive to the observer's computer (left to right). The heart of the design is the "ODAP backend," a daemon that manages authorization and bundling of ingested files. It monitors the ODAP queue for newly ingested data, verifies the user has access, queries the KOA database for all raw and reduced files for each program that have been ingested on that date, and then forwards the metadata to the Flask server, which in turn forwards metadata to the front-end client that displays the metadata. This operation is shown at the top of the diagram. 

Two other operations are shown in Figure 2: requesting a file and requesting a wget script. To request a file, the ODAP Backend detects a file request, verifies the user has access to it, locates the file, reads the contents into a bytes object, and then ships it to the client. A request for a wget file is more complex than for a single file: the backend detects a request for a wget file, queries the database for all raw and reduced files for each program that have been ingested on UTDATE, sends queries to the database replicate filters placed on the table, assembles all rows into a wget script, and then reads the wget script content into a bytes object that is sent to the client.

\articlefigure{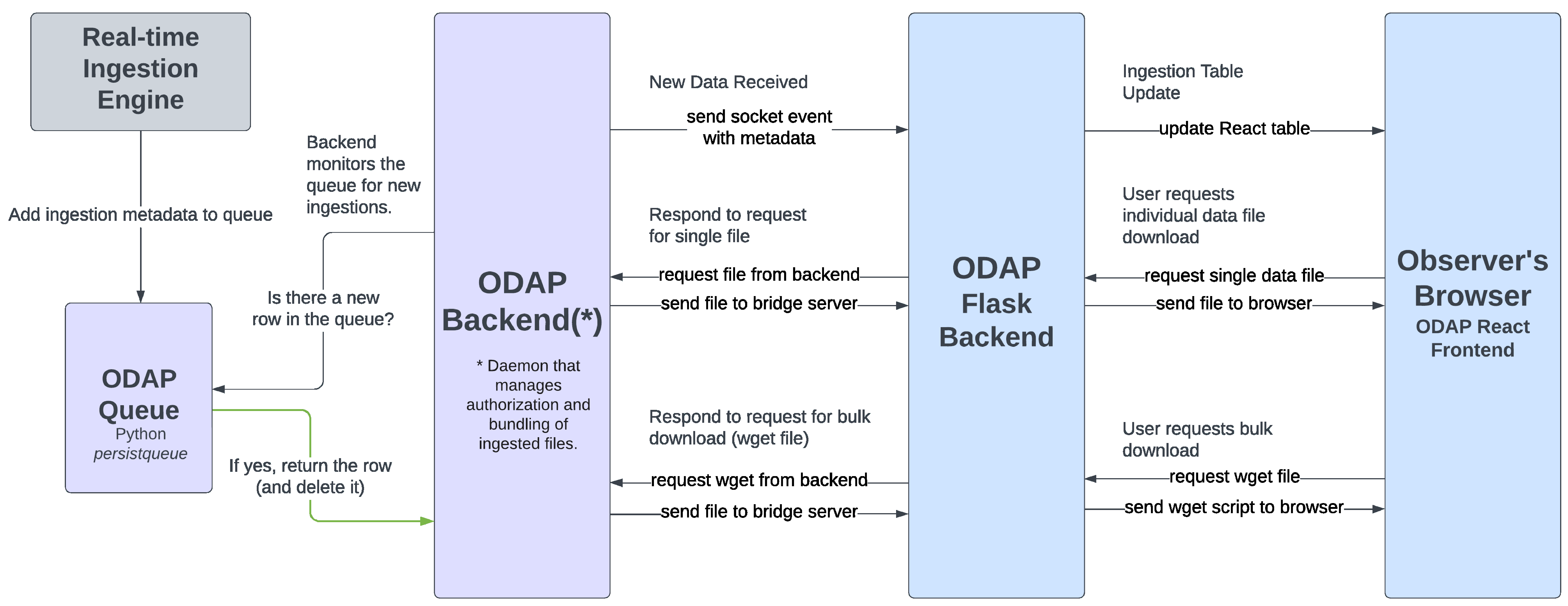}{ex_fig2}{The Design of the ODAP}

For simplicity, the authentication process is not shown in Figure 2. Authentication in the ODAP is provided by a JSON web token (JWT), an open standard (RFC 7519) that defines a compact, self-contained way for securely transmitting information between parties as a JSON object. \footnote{\url{https://auth0.com/docs/secure/tokens/json-web-tokens}}

The authentication process operates as follows:

\begin{itemize}
    
\item The authentication library verifies a userid and password against entries in the KOA database, from which it acquires a list of all Semester IDs authorized to that user. (Semester IDs are a combination of observing semester and program ID.) An https call to WMKO's nightly schedule cross-matches the result with the observer's authorized semester IDs, which are bundled with user info (e.g., name, koaid, keckid, and if available, email) and hashed into an access$-$token.
\item The authentication library returns an access$-$token, refresh$-$token, and some additional user data, such as the user's email, authorized semids, etc.
\item The tokens contain the authorized semids and cannot be altered without re-initiating the login process.
\item The tokens are sent with every subsequent call to the RTI backend, where it is deciphered and used to obtain accurate and user access information that cannot be tampered.
\end{itemize}

\acknowledgments  The Keck Observatory Archive (KOA) is a collaboration between the NASA Exoplanet Science Institute (NExScI) and the W. M. Keck Observatory (WMKO). NExScI is sponsored by NASA's Exoplanet Exploration Program and operated by the California Institute of Technology in coordination with the Jet Propulsion Laboratory (JPL). 

The observatory was made possible by the generous financial support of the W. M. Keck Foundation. The authors wish to recognize and acknowledge the very significant cultural role and reverence that the summit of Mauna Kea has always had within the indigenous Hawaiian community. We are most fortunate to have the opportunity to conduct observations from this mountain.

\bibliography{P616}  


\end{document}